\documentclass[aps,prb,reprint,amsmath,amssymb,superscriptaddress,floatfix]{revtex4-2}
\usepackage{graphicx}
\usepackage[colorlinks, citecolor=blue, linkcolor=blue]{hyperref}
\usepackage{txfonts}
\usepackage{bm}		% Bold math

\newcommand{\CNN}{Centre de Nanosciences et de Nanotechnologies, CNRS, Universit\'e Paris-Saclay, 91120 Palaiseau, France}
\newcommand{\Napoli}{Department of Electrical Engineering and ICT, Universit{\`a} degli Studi di Napoli Federico II, 80125 Napoli, Italy}

\begin{document}

\title{Mode-resolved micromagnetics study of parametric spin wave excitation in thin-film disks}

\author{Maryam Massouras}
\affiliation{\CNN}
\author{Salvatore Perna}
\author{Massimiliano d'Aquino}
\author{Claudio Serpico}
\affiliation{\Napoli}
\author{Joo-Von Kim}
\email{joo-von.kim@c2n.upsaclay.fr}
\affiliation{\CNN}
\date{August 28, 2024}

\begin{abstract}
We present a computational study of the parametric excitation of spin waves in thin film disks with a mode-resolved approach. The method involves projecting out the time-dependent magnetization, computed using micromagnetics simulations, onto the spatial profile of the eigenmodes that are obtained from the linearization of the equations of motion. Unlike spectral analysis in the frequency domain, the projection allows for the analysis of transient mode dynamics under parametric excitation. We apply this method to parallel pumping of quantized spin wave modes in in-plane magnetized thin-film disks, where phenomena such as frequency pulling, mutual phase locking, and higher-order magnon scattering processes are identified.
\end{abstract}

\maketitle

\section{Introduction}
Spin waves in patterned magnetic thin films differ from their counterparts in bulk systems in many important ways. For example, the finite-sized nature of a confined system such as a thin-film disk results in spatially nonuniform dipolar fields and pinning at boundary edges that play a crucial role in governing the spin wave eigenmode spectrum~\cite{Guslienko:2000vn, Demokritov:2001kk, Jorzick:2001il, Jorzick:2002fo, Bayer:2003gu, Gubbiotti:2003fp, Guslienko:2003jv, Bayer:2005gl, Guslienko:2005jy, Zivieri:2006ct, Demidov:2008jd, Neudecker:2008kj, Shaw:2009gr, Nembach:2011bd, Guo:2013fk, GarciaSanchez:2014dw, Dutra:2019cj, Zingsem:2019gt, Porwal:2019hl, Mondal:2020fn, Nembach:2021co, Perna:2022nm, Kharlan:2022mo}. Nonlinear processes such as parametric excitation and magnon-magnon scattering~\cite{Sparks:1964, Lvov:1994, Gurevich:1996, Bertotti:2009} are also greatly influenced by finite-size effects, where by virtue of a discrete mode spectrum certain scattering processes may be inhibited or enhanced~\cite{Ulrichs:2011dz, Demidov:2011iv, Camley:2014cg}. From a theoretical and computational perspective, it is therefore interesting to examine how parametrically excited spin wave eigenmodes interact with one another in arbitrary confined geometries, for which analytical descriptions of the eigenmodes may be lacking.

Such considerations become particularly acute for potential applications~\cite{Barman.2021tm, Pirro:2021ai, Chumak:2022ai}, where detailed knowledge of nonlinear spin wave processes is important for tasks in information processing. For example, the parametric excitation of spin waves in waveguides~\cite{Bracher:2013ej} or narrow conduits~\cite{Hwang.2021pe, Heinz:2022pg} is crucial for populating certain spin wave modes, which can also shed light on subsidiary processes such as spin pumping driven by magnon decay~\cite{Hahn.2021eo}. For tasks such as pattern recognition, a recent study by K{\"o}rber \emph{et al.} showed that nonlinear spin wave interactions can be exploited within the paradigm of physical reservoir computing~\cite{Korber:2023pr}. The approach involves mapping a recurrent neural network to a dynamical system~\cite{Maass:2002kf, Tanaka:2019ra}, whereby input signals are transformed by the nonlinear dynamics of the physical system, and machine learning is performed on the outputs to perform tasks such as classification. In contrast to other proposals for using spin waves in reservoir computing, which rely on spatial nodes (nonlinear interference patterns~\cite{Nakane:2018jc, Ichimura.2021an} or rf fingerprinting in artificial spin ice~\cite{Gartside:2022rt}) or temporal nodes (delayed feedback~\cite{Watt:2020di, Watt:2021ia, Nikitin:2022tn, Watt:2023lt}), the approach in Ref.~\onlinecite{Korber:2023pr} maps the node structure into reciprocal space using the spin wave eigenmodes of a vortex state in a thin-film ferromagnetic disk, where inputs correspond to radial eigenmodes that couple to an external radio-frequency (rf) field, while output modes correspond to azimuthal eigenmodes that are driven by three-magnon scattering processes. Understanding the nonlinear magnon processes in nanostructured materials is therefore crucial if one seeks to formulate computational tasks based on such phenomena.

To this end, we revisit the problem of parametric spin wave excitation via parallel pumping in confined magnetic systems, specifically an in-plane magnetized disk for which recent work have shown rich behavior in the mode spectrum~\cite{Ulrichs:2011dz, Guo:2014cw, Srivastava:2023io}. We present a mode-resolved approach in computational micromagnetics, which allows us to compute the transient dynamics of the mode amplitude and intensities of excited and scattered modes. This method differs from the more conventional approaches based on analyses of the power spectral density in the frequency domain, since it allows nonlinear phenomena such as frequency pulling, mutual synchronization of excited spin wave modes, and temporal dynamics such as mode growth and inhibition to be studied in detail.

This paper is organized as follows. In Sec. II, we describe the system studied and the methodology behind the mode-filtering technique. In Sec. III, we apply this technique to the problem of the parametric excitation via parallel pumping of the quasi-uniform mode. Section IV discusses parametric thresholds and their functional form for thin film disks, which is followed by Sec. V in which mode generation at high power is discussed. Finally, we present a discussion and concluding remarks in Sec. VI.

%%
%	Section: Geometry and Method
%%
\section{Geometry and method}
\subsection{Geometry}
\begin{figure}
	\centering\includegraphics[width=7cm]{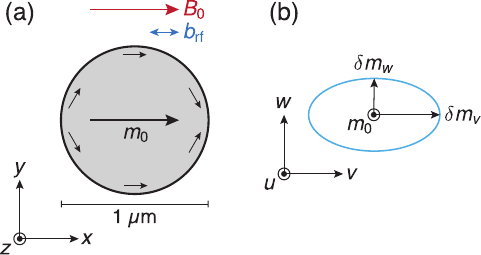}
	\caption{(a) Geometry of the thin-film disk studied, where we assume the static $B_0$ and rf driving field $b_\mathrm{rf}$ are applied along the $x$ direction. The static magnetization $m_0$ is largely oriented along the direction of $B_0$, but slight deviations appear near the boundary edges, which arise from minimizing dipolar surface charges. (b) Schematic of the steady-state elliptical precession of magnetization, where it is assumed that the static orientation lies along the $u$ axis, while the dynamic components of the magnetization span the $vw$ plane.}
	\label{fig:geometry}
\end{figure}

The system studied is illustrated in Fig.~\ref{fig:geometry}(a). 
%%%
%%%
We consider a ferromagnetic thin film disk with a diameter of 1 $\mu$m and a thickness of 50 nm, where a static magnetic field, $B_0$, and a sinusoidal radio-frequency (rf) field, $b_\mathrm{rf}$, are both applied along the $x$ direction, which corresponds to the parallel pumping configuration. To excite a normal mode in this configuration, it is necessary to apply an rf field at twice the normal mode frequency. $\mathbf{m}_0$ denotes the static configuration of the magnetization in the disk, which is largely aligned along the $x$ direction as a result of $B_0$, but non-uniformities can appear at the edges through the minimization of surface dipolar charges. In Fig.~\ref{fig:geometry}(b), we present a local coordinate system in which $\mathbf{m}_0$ is aligned along $u$, while $v,w$ denote the two components perpendicular to $u$ along which magnetization fluctuations ($\delta m_{v,w}$) lie. We will make use of this coordinate transformation later.

The time evolution of the magnetization is calculated using the \textsc{MuMax3} code~\cite{Vansteenkiste:2014et}, which employs the finite-difference method to perform the numerical time integration of the Landau-Lifshitz equation with Gilbert damping,
\begin{equation}
	\frac{d \mathbf{m} }{dt} = -|\gamma_0|  \mathbf{m} \times \left( \mathbf{H_{\mathrm{eff}}} + \mathbf{h_{\mathrm{th}}} \right) + \alpha \mathbf{m} \times \frac{d \mathbf{m} }{dt},
	\label{eq:LLG}
\end{equation}
where $\mathbf{m}(\mathbf{r},t)$ is the unit vector representing the magnetization, $\gamma_0 = \mu_0 g \mu_B / \hbar$ is the gyromagnetic constant, and $\alpha$ is the Gilbert damping constant. The effective field, $\mathbf{H}_\mathrm{eff} = -(1/\mu_0M_s)\delta U/\delta \mathbf{m}$, represents a variational derivative of the total magnetic energy $U$ with respect to the magnetization (with $M_s$ being the saturation magnetization), where $U$ contains contributions from the Zeeman, nearest-neighbor Heisenberg exchange, and dipole-dipole interactions. Finite-temperature effects are modeled by including a random field $\mathbf{h}_\mathrm{th}$ in the effective field, which has zero mean, $\langle  \mathbf{h}_\mathrm{th} \rangle = 0$, and represents a Gaussian white noise with the spectral properties
\begin{equation}
\langle  h_{\mathrm{th},i}(\mathbf{r},t) h_{\mathrm{th},j}(\mathbf{r}',t') \rangle = \frac{2 \alpha k_B T}{\mu_0 V} \delta_{ij} \delta(\mathbf{r}-\mathbf{r}') \delta(t-t'),
\end{equation}
where $i,j$ represent the different Cartesian components of the field vector, and $V$ is the volume of the unit cell~\cite{Brown:1963cb}. Time integration of the Langevin problem is performed using an adaptive time-step scheme~\cite{Leliaert:2017ci}.

In all simulations presented here, the 1 $\mu$m-diameter, 50 nm-thick disk was discretized using $256 \times 256 \times 1$ finite difference cells. We used an exchange constant of $A = 3.7$ pJ/m, a saturation magnetization of $M_s = 141$ kA/m, and a Gilbert damping constant of $\alpha = 1.5\times 10^{-3}$, which are modeled after experiments on thin film, $\mu$m-sized disks of yttrium iron garnet~\cite{Srivastava:2023io}.

%%
%	Section: Supercell
%%
\subsection{Supercell method for spectral analysis}
\begin{figure}
	\centering\includegraphics[width=8cm]{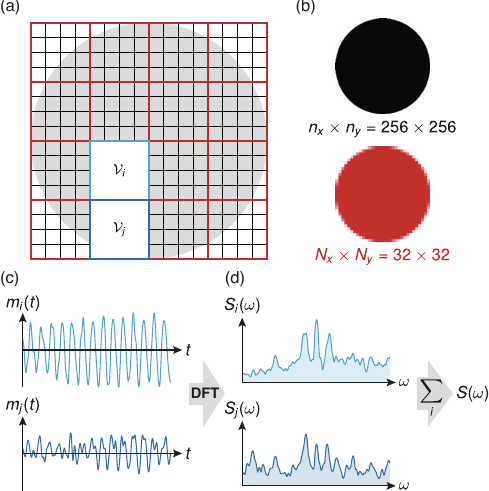}
	\caption{(a) Supercells, delimited by red lines, are constructed from ensembles of finite-difference cells, delimited by black lines. (b) Finite-difference and supercell discretization used for the 1 $\mu$m disk. (c) Time evolution of magnetizations $m_{i,j}$ in the supercells $\mathcal{V}_{i,j}$ shown in (a). (d) Power spectrum $S_{i,j}(\omega)$, shown on a log scale, of the magnetization in $\mathcal{V}_{i,j}$ calculated from the discrete Fourier transform of the amplitudes in (c). The overall power spectrum is obtained by summing over the individual mode spectra.}
	\label{fig:supercell}
\end{figure}

Two methods to estimate the power spectral density of excitations, $S(\omega)$, are considered here. The first is the supercell approach, which involves averaging the dynamics over ensembles of finite-difference cells as shown in Fig.~\ref{fig:supercell}(a). 
%%%
%%%
In this schematic illustration, the finite-difference cells are delimited by black lines and represent the ultimate spatial resolution of the system studied, while the supercells delimited by red lines represent a coarse-grained version of the simulated dynamics. Note that this only applies to the simulation output; Eq.~(\ref{eq:LLG}) is always solved for each finite-difference cell, while the choice of the supercell size is based on the tradeoff between accuracy and convenience. For the examples discussed further below, the supercells used correspond to blocks of $8 \times 8 \times 1$ finite-difference cells, leading to a coarse-grained output of $32 \times 32 \times 1$ supercells as shown in Fig.~\ref{fig:supercell}(b).

The time-dependent magnetization is averaged within each supercell, $\mathcal{V}_i$, from which we can project out $\delta m_{i,v}$ and $\delta m_{j,v}$, the fluctuations transverse to the static magnetization, $\mathbf{m}_0$, as shown in Fig.~\ref{fig:supercell}(c). The power spectrum of these transverse fluctuations, $S_i(\omega)$, is then obtained from the discrete Fourier transform (DFT) of $\delta m_{i}$. The total power spectrum of the spin wave excitations is then obtained by summing over the contributions from each supercell, $S(\omega) = \sum_i S_i(\omega)$. One important caveat should be noted here. By virtue of averaging the magnetization within each supercell, we lose information about higher-frequency spin waves whose wavelengths are smaller than the supercell size. The choice of the supercell size is therefore also guided by the frequency regime under study.

From a technical perspective, the supercell approach offers a computationally efficient method for obtaining $S(\omega)$. A more direct approach would involve recording the full magnetization state at each desired time step, from which $S(\omega)$ is calculated from the time-dependent magnetization in each finite-difference cell~\cite{Mcmichael:2005co}, rather than in each supercell. While this allows for more accurate spatial profiles to be obtained and avoids the high-frequency cutoff, it requires greater computational resources in the post-processing phase of the data analysis. With the supercell approach, on the other hand, the averaged magnetization in each supercell is computed on the fly (e.g., by using the \texttt{Crop} function in \textsc{MuMax3}) and exported directly during the course of the simulation run, which greatly reduces the need for post-processing. Its inclusion in the discussion here is simply to provide a reference for a spatially resolved approach against which we compare the mode-filtering method, which we discuss next.

%%
%	Section: Mode filtering
%%
\subsection{Mode-filtering method for spectral analysis}
\begin{figure}
	\centering\includegraphics[width=8cm]{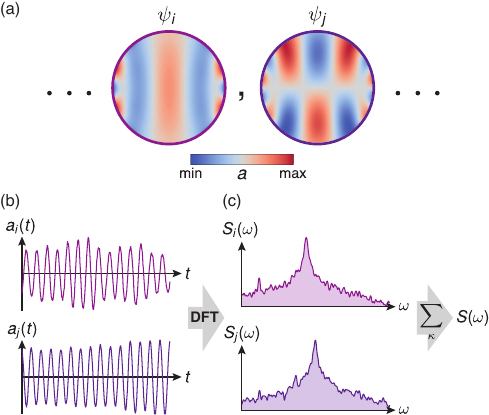}
	\caption{(a) Examples of spatial profiles $\psi(\mathbf{r})$. (b) Mode amplitude $a(t)$ as a function of time. (c) Power spectrum $S(\omega)$, shown on a log scale, of the modes in (a) calculated from the discrete Fourier transform of the amplitudes in (b). The overall power spectrum is obtained by summing over the individual mode spectra.}
	\label{fig:modeproj}
\end{figure}

The second method we use for spectral analysis is the mode-filtering method, which is the focus of this paper and schematized in Fig.~\ref{fig:modeproj}. 
%%%
%%%
The starting point is the determination of the magnetization normal oscillation modes associated to the confined structure under investigation. Initially, we assume that $\mathbf{m}(\mathbf{r}, t) = \mathbf{m}_0(\mathbf{r}) + \delta \mathbf{m}(\mathbf{r}, t)$ in Eq.~(\ref{eq:LLG}) and we analyze the system in the regime of small $\delta \mathbf{m}(\mathbf{r}, t)$, where its dynamics are governed by a linearized equation of motion. Diagonalizing this linear equation yields the normal modes, denoted as $\bm{\varphi}_\kappa(\mathbf{r})$, which are complex-valued vector fields transverse to $\mathbf{m}_0(\mathbf{r})$~\cite{dAquino:2009ct, perna:2022cm}. Specifically, $\delta \mathbf{m}(\mathbf{r}, t)$ can be expressed in terms of these normal modes as
\begin{equation}
\delta \mathbf{m}(\mathbf{r}, t) = \sum_{\kappa=1}^{\infty} c_\kappa(t) \bm{\varphi}_\kappa(\mathbf{r}) + \mathrm{c.c.},
\label{eq:normalmodes}
\end{equation}
where ``c.c.'' means complex conjugate, $\kappa$ represents the mode index, and $\bm{\varphi}_\kappa(\bm{r})$ is given by
\begin{equation}
\bm{\varphi}_\kappa(\mathbf{r}) = \varphi_{\kappa,v}(\mathbf{r}) \, \mathbf{e}_v(\mathbf{r}) + \varphi_{\kappa,w}(\mathbf{r}) \, \mathbf{e}_w(\mathbf{r}),
\end{equation}
where $\mathbf{e}_v(\mathbf{r}), \mathbf{e}_w(\mathbf{r})$ are Cartesian unit vectors spanning the plane orthogonal to $\mathbf{m}_0(\mathbf{r})$ [Fig.~\ref{fig:geometry}(b)], and $\varphi_{\kappa,v}$ and $\varphi_{\kappa,w}$ are the components of $\bm{\varphi}_h(\mathbf{r})$ along $\mathbf{e}_v(\mathbf{r})$ and $\mathbf{e}_w(\mathbf{r})$, respectively.

In contrast, in conventional spin wave analysis for a continuous medium, the basis functions are plane waves,
\begin{equation}
\bm{\varphi}_\mathbf{k}(\mathbf{r}) = \bm{\phi}_\mathbf{k} \, e^{\pm i \, \mathbf{k} \cdot \mathbf{r}}, 
\end{equation}
where $\mathbf{k}$ is the wave vector and $\bm{\phi}_\mathbf{k}$ a constant vector with complex entries, orthogonal to the direction of the magnetization ground state, which describes the polarization of the spin wave with wave vector $\mathbf{k}$. Consequently, $\delta \mathbf{m}(\mathbf{r}, t)$ can be expanded in spin waves as
\begin{equation}
\delta \mathbf{m}(\mathbf{r}, t) = \sum_{\mathbf{k}} \mathbf{a}_\mathbf{k}(t) e^{i \, \mathbf{k} \cdot \mathbf{r}} + \mathrm{c.c.},
\label{eq:spinwaves}
\end{equation}
where $\mathbf{a}_\mathbf{k}(t) = a_\mathbf{k}(t) \bm{\phi}_\mathbf{k}$ are time-dependent vectors and $a_\mathbf{k}(t)$ are appropriate complex expansion coefficients.

It is important to notice that the $\mathbf{r}$ dependence of the spin wave modes is contained in the scalar factor $e^{\pm i \, \mathbf{k} \cdot \mathbf{r}}$, while in the normal mode expansion [Eq.~(\ref{eq:normalmodes})] the $\mathbf{r}$ dependence is taken into account by the vector fields $\bm{\varphi}_\kappa(\mathbf{r})$. To facilitate the comparison of the description based on the normal modes with spin wave analysis, and subsequently relate expansions (\ref{eq:spinwaves}) and (\ref{eq:normalmodes}), we introduce approximately orthogonal scalar basis functions
\begin{equation}
\psi_\kappa(\mathbf{r}) = \mathrm{Re}\left[ g_{\kappa,v} \varphi_{\kappa,v} + g_{\kappa,w} \varphi_{\kappa,w} \right],
\end{equation}
where $g_{\kappa,v},g_{\kappa,w}$ are normalization constants ensuring
\begin{equation}
\frac{1}{V} \int_V \left[ \psi_\kappa(\mathbf{r}) \right]^2 dV = 1.
\end{equation}
These basis functions are approximately orthogonal as the overlap integral
\begin{equation}
F_{\kappa,\kappa'} = \frac{1}{V} \int_V \psi_\kappa(\mathbf{r}) \, \psi_{\kappa'}(\mathbf{r}) \;  dV
\end{equation}
slightly differs from the Kronecker delta, $\delta_{\kappa,\kappa'}$. For the problem at hand, the off-diagonal elements of $F_{\kappa,\kappa'}$ are less than $10^{-2}$~\cite{SM}.

We refer to the scalar fields $\psi_\kappa(\mathbf{r})$ as the spatial mode profiles [Fig.~\ref{fig:modeproj}(a)] and express $\delta \mathbf{m}(\mathbf{r}, t)$ in terms of these as
\begin{equation}
\delta \mathbf{m}(\mathbf{r}, t) = \sum_{\kappa=1}^{\infty} \mathbf{a}_\kappa(\mathbf{r},t) \psi_\kappa(\mathbf{r}),
\label{eq:modefilter}
\end{equation}
where
\begin{equation}
\mathbf{a}_\kappa(\mathbf{r},t) = a_{\kappa,v}(t) \mathbf{e}_v(\mathbf{r}) + a_{\kappa,w}(t) \mathbf{e}_w(\mathbf{r}).
\end{equation}
We can therefore interpret $\psi_\kappa(\mathbf{r})$ as the modified version of the scalar factor present in the usual spin wave expression~(\ref{eq:spinwaves}), with $a_{\kappa,v}(t), a_{\kappa,w}(t)$ corresponding to the components of the complex mode amplitude $\mathbf{a}_\mathbf{k}(t)$. As such, Eq.~(\ref{eq:modefilter}) can be viewed as a generalization of the traditional spin wave expansion applied to confined structures.

Equation~\ref{eq:modefilter} is the basis of the mode-filtering method. The use of scalar mode profile $\psi_\kappa(\mathbf{r})$ functions enables a more direct comparison between the mode-filtering method and traditional spin wave analysis that will be discussed later in the paper. In contrast to the supercell case [Fig.~\ref{fig:supercell}(c)], where the power spectrum of the individual supercells contains information on all the magnetization fluctuations in the system [Fig.~\ref{fig:supercell}(d)], the mode amplitudes $a_{\kappa,v}(t)$ and $a_{\kappa,w}(t)$ only contain information about the associated mode profile $\psi_\kappa(\mathbf{r})$ [Fig.~\ref{fig:modeproj}(b)].

%%
%	Parametric excitation of the uniform mode
%%
\section{Parametric excitation of the uniform mode}
\label{sec:paramexc}
In this section, we focus on the parametric excitation of the quasi-uniform mode of the 1-$\mu$m disk studied to illustrate the salient features of the mode-filtering technique, in particular, in how it can overcome shortcomings of the supercell approach (and, by extension, any spatially resolved frequency-domain analysis).

\subsection{Spectral analysis with supercells}
\begin{figure}
	\centering\includegraphics[width=8cm]{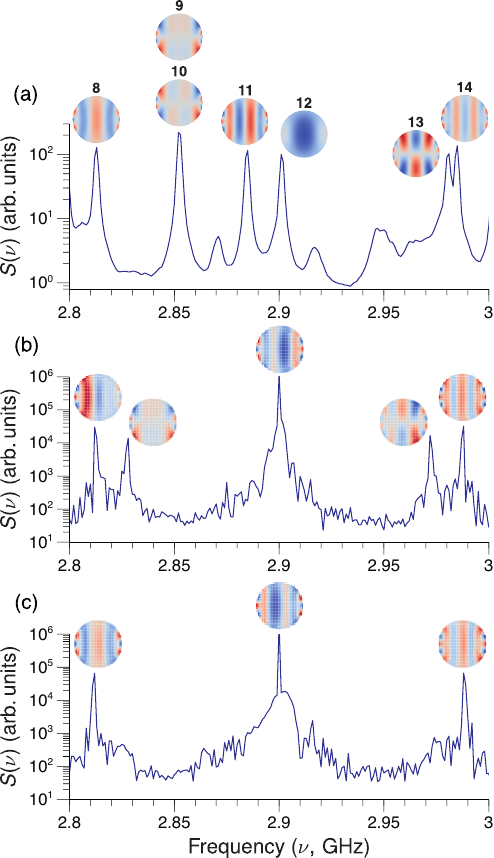}
	\caption{Power spectral density $S(\nu)$ computed with the supercell method of spin waves around $\nu = \nu_{12} = 2.901$ GHz at $T=300$ K. (a) Thermal spectra with peaks labeled by normal mode indices and the corresponding spatial profiles obtained from linearizing the equations of motion.	(b) Parametric excitation at $\nu_\mathrm{rf}= 2 \nu_{12} = 5.802$ GHz with $b_\mathrm{rf} = 1.0$ mT. The spatial profiles represent the back-transformed data from the supercell spectra. (c) Parametric excitation at 5.802 GHz and 1.5 mT. The spatial profiles represent the back-transformed data from the supercell spectra}
	\label{fig:psd_supercell}
\end{figure}

We first consider the thermal spectrum of spin wave excitations at 300 K, which is shown in Fig.~\ref{fig:psd_supercell}(a).
%%%
%%%
The stochastic dynamics under a static field $B_0 = 50$ mT is simulated over an interval of 100 $\mu$s, and subsequently the Welch method~\cite{Welch:1967} with half-overlapping 1-$\mu$s Hann windows is applied to obtain the averaged power spectrum. For this value of $B_0$, the frequency of the quasi-uniform mode ($\psi_{12}$) is $\nu_{12} = 2.901$ GHz, so we restrict the plot in Fig.~\ref{fig:psd_supercell} to a frequency range of interest about this mode frequency. In the inset above Fig.~\ref{fig:psd_supercell}(a), the spatial profiles of selected modes within this frequency range are shown, where the indices $\kappa$ are ordered according to the mode frequency. The quasi-uniform mode is not the lowest-frequency mode, as there are a number of edge modes and modes with a backward-volume character, e.g., $\kappa=8$ and $\kappa=11$, which are lower in frequency. We note that the profiles in Fig.~\ref{fig:psd_supercell}(a) are those computed with the linearization technique~\cite{dAquino:2009ct, perna:2022cm}, which are matched with the peaks in the thermal PSD computed with the supercell technique.

The power spectrum corresponding to a parametric excitation of $b_\mathrm{rf} = 1.0$ mT at $\nu_\mathrm{rf} = 2 \nu_{12} = 5.902$ GHz, applied over an interval of 1 $\mu$s, is shown in Fig.~\ref{fig:psd_supercell}(b). 
The top inset shows spatial mode profiles computed by back Fourier transform of the supercell data. While it is unsurprising that the dominant response is found at $\nu_\mathrm{rf}/2 = 2.901$ GHz, the spatial profile corresponding to this frequency does not resemble the quasi-uniform mode ($\psi_{12}$), but rather a distorted version of the neighboring $\psi_{11}$ mode. Along with the primary excitation, we can observe two pairs of satellite peaks centered about $\nu_\mathrm{rf}/2$. Frequency pulling in some of these satellites peaks can be seen, where $\psi_{9}$ and $\psi_{13}$ exhibit a frequency red shift while $\psi_{14}$ experiences a slight frequency blue shift. Like with the primary excitation, modes $\psi_8$ and $\psi_{13}$ appear distorted with respect to their linearized profiles. As the driving field amplitude is increased to $b_\mathrm{rf} = 1.5$ mT, one pair of the satellite peaks disappears ($\psi_9$, $\psi_{13}$), while the the other pair ($\psi_8$, $\psi_{14}$) remains at the same frequency splitting as under $b_\mathrm{rf} = 1.0$ mT.

\subsection{Spectral analysis with mode filtering}
\begin{figure}
	\centering\includegraphics[width=8cm]{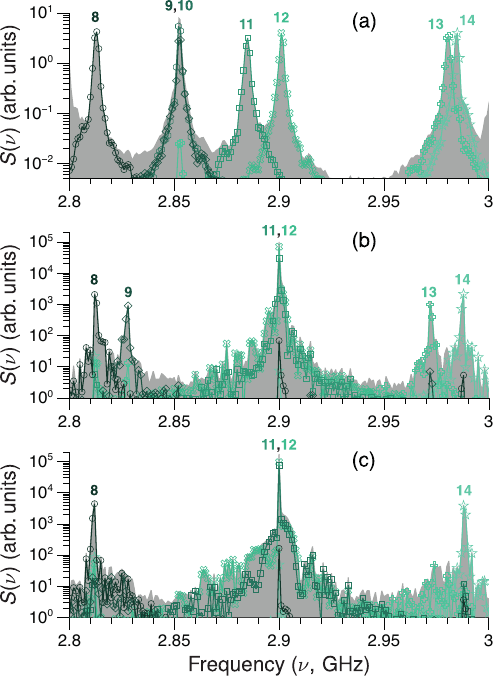}
	\caption{Power spectral density $S(\nu)$ computed with the mode-filtering method of spin waves around $\nu = \nu_{12} = 2.901$ GHz at $T = 300$ K. The spectrum of each mode $S_\kappa(\nu)$ is presented in a different color, while the shaded, gray curve in the background represents the total PSD $S(\nu)$. (a) Thermal spectra. (b) Parametric excitation at $\nu_\mathrm{rf}= 2 \nu_{12}= 5.802$ GHz with 1.0 mT. (c) Parametric excitation at 5.802 GHz and 1.5 mT. The peaks are indexed by the normal mode number $\kappa$.}
	\label{fig:psd_modeproj}
\end{figure}

We can shed light on the supercell results above using the mode-filtering method, with which we used to study the same cases as shown in Fig.~\ref{fig:psd_supercell}. These results are presented in Fig.~\ref{fig:psd_modeproj}.
%%%
%%%
The thermal power spectrum is shown in Fig.~\ref{fig:psd_modeproj}(a), where the gray background represents the full $S(\nu)$ while the mode-resolved responses are shown in color for selected modes. In contrast to the supercell case, we find that a 10-$\mu$s simulation with half-overlapping 1-$\mu$s windows for the Welch method was sufficient, as the projection effectively filters out contributions from other spin wave modes. The peaks are numbered according to the spatial profiles given in Fig.~\ref{fig:psd_supercell}(a), which provide an important verification that the same modes are identified using the two methods.

The power spectrum for the parametric excitation at $b_\mathrm{rf} = 1.0$ mT is shown in Fig.~\ref{fig:psd_modeproj}(b). The mode filtering reveals that the primary excitation corresponds to modes $\psi_{11}$ and $\psi_{12}$ being excited simultaneously, with $\psi_{12}$ more strongly excited than $\psi_{11}$ but comparable in intensity. This indicates that the deformed profile seen in Fig.~\ref{fig:psd_supercell}(b) results from a superposition of $\psi_{11}$ and $\psi_{12}$. The frequency pulling associated with modes $\psi_{9}$, $\psi_{11}$, $\psi_{13}$, and $\psi_{14}$ can be seen by inspection upon comparing Fig.~\ref{fig:psd_modeproj}(a) and Fig.~\ref{fig:psd_modeproj}(b). However, no other mode appears to be associated with mode $\psi_8$, which suggests that the distorted profile observed for the lowest-frequency satellite peak in Fig.~\ref{fig:psd_supercell}(b) is not related to the coexistence of several modes. The coexistence of modes $\psi_{11}$ and $\psi_{12}$ persist under the stronger driving field of $b_\mathrm{rf} = 1.5$ mT, as shown in Fig.~\ref{fig:psd_supercell}(c), with only the $\psi_8$ and $\psi_{14}$ satellites present as discussed previously.

\begin{figure}
	\centering\includegraphics[width=8cm]{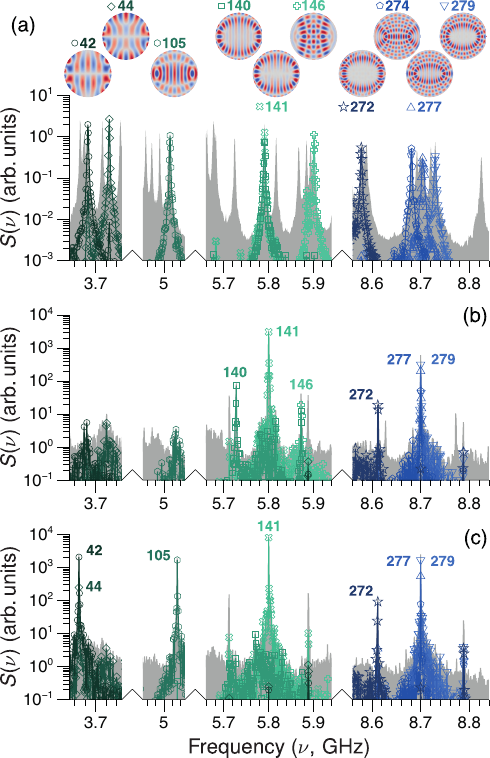}
	\caption{Power spectral density $S(\nu)$ computed with the mode-filtering method of spin waves around $3\nu_{12}/2$, $2 \nu_{12}$, and $3 \nu_{12}$ at $T = 300$ K. The spectrum of each mode $S_\kappa(\nu)$ is presented in a different color, while the shaded, gray curve in the background represents the total PSD $S(\nu)$.  (a) Thermal spectra. (b) Parametric excitation at $\nu_\mathrm{rf}=2\nu_{12}=5.802$ GHz with $b_\mathrm{rf} = 1.0$ mT. (c) Parametric excitation at 5.802 GHz and 1.5 mT. The peaks are indexed by the normal mode number $\kappa$, with the corresponding spatial profiles shown in the top inset of (a).}
	\label{fig:psd_modeproj_otherf}
\end{figure}

Similar frequency pulling, satellite peaks, and mode coexistence are also observed about other (integer fraction) multiples of $\nu_{12}$, such as about $\nu = 3 \nu_{12} / 2$, $2 \nu_{12}$, and $3\nu_{12}$ as shown in Fig.~\ref{fig:psd_modeproj_otherf}.
%%%
%%%
The thermal spectrum is presented in Fig.~\ref{fig:psd_modeproj_otherf}(a), where some modes of interest are highlighted and their spatial profiles are shown in the inset. Under $b_\mathrm{rf} = 1.0$ mT, the close frequency degeneracy for modes $\psi_{140}$ and $\psi_{141}$ is lifted, with $\psi_{141}$ locking to the frequency $\nu_\mathrm{rf} = 2\nu_{12}$ and $\psi_{140}$ and $\psi_{146}$ becoming satellite peaks. Around $3\nu_{12}$, on the other hand, we observe that modes $\psi_{277}$ and $\psi_{279}$, which possess distinct spectral peaks in the pure thermal regime, lock onto $\nu = 3\nu_{12}$ under parametric excitation at $\nu_\mathrm{rf} = 2 \nu_{12}$ and is accompanied by $\psi_{272}$ which undergoes a frequency blueshift. As the driving field amplitude is increased to $b_\mathrm{rf} = 1.5$ mT, satellite peaks appear about $\nu = 3\nu_{12}/2$, where the lower frequency satellite comprises modes $\psi_{42}$ and $\psi_{44}$, while the higher frequency satellite corresponds to mode $\psi_{105}$.

\subsection{Phase portraits}
\begin{figure}
	\centering\includegraphics[width=7cm]{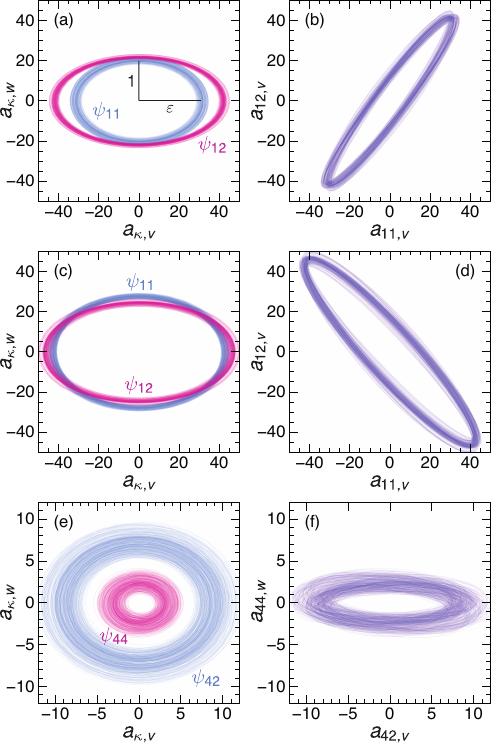}
	\caption{Phase portraits (a), (c), (e) and Lissajous curves (b), (d), (f) for parametric excitation with $\nu_\mathrm{rf} = 2\nu_{12} = 5.802$ GHz and $B_0 = 50$ mT. (a) $a_{11},a_{12}$ and (b) $(a_{11,v},a_{12,v})$ under under $b_\mathrm{rf} = 1.0$ mT. (c) $a_{11},a_{12}$ and (d) $(a_{11,v},a_{12,v})$ under under $b_\mathrm{rf} = 1.5$ mT. (e) $a_{42},a_{44}$ and (f) $(a_{42,w},a_{44,w})$ under under $b_\mathrm{rf} = 1.5$ mT. The curves are drawn from the last 100-ns segment of a 1-$\mu$s simulation.}
	\label{fig:lissajous}
\end{figure}

To better understand the coupled dynamics of $\psi_{11}$ and $\psi_{12}$, which are the primary excitations under parallel pumping at $\nu_\mathrm{rf} = 2 \nu_{12} = 5.902$ GHz, we examine their phase portraits based on the mode amplitudes $a_{\kappa,v}$ and $a_{\kappa,w}$. Figure~\ref{fig:lissajous}(a) illustrates the phase space dynamics over the last 100 ns of the 1-$\mu$s long simulation, which was performed to obtain the power spectral density in Figs.~\ref{fig:psd_modeproj} and ~\ref{fig:psd_modeproj_otherf}.
%%%
%%%
The phase portraits illustrate a well-defined limit cycle for each of $\psi_{11}$ and $\psi_{12}$, where the ellipticity of the limit cycle reflects the average ellipticity $\varepsilon$ of the spin precession associated with the mode, as illustrated for mode $\psi_{11}$ in Fig.~\ref{fig:lissajous}(a). The finite width of the limit cycles is due to thermal fluctuations, which result in both phase and amplitude noise. The mutual dynamics of modes $\psi_{11}$ and $\psi_{12}$ can be seen in Fig.~\ref{fig:lissajous}(b), which features a Lissajous curve constructed from the $v$ component of $a_{11}$ and $a_{12}$. The resulting ellipse here indicates that the modes are in fact mutually phase-locked, with a constant phase difference of approximately $14.7^\circ$. Under the larger excitation field of $b_\mathrm{rf} = 1.5$ mT, the limit cycles increase in size [Fig.~\ref{fig:lissajous}(c)] and mutual phase-locking persists [Fig.~\ref{fig:lissajous}(d)], but with a phase difference of approximately $-160.9^{\circ}$ as reflected by the change in orientation of the Lissajous ellipse.

Similar phase-locking behavior is observed for the modes $\psi_{42}$ and $\psi_{44}$, which appear as secondary excitations under $b_\mathrm{rf} = 1.5$ mT as shown in Figs.~\ref{fig:psd_modeproj_otherf}(c) and \ref{fig:transients}(h). The natural frequencies of these modes are $\nu_{42} = 3.685$ GHz and $\nu_{44} = 3.726$ GHz, respectively, but become phase-locked at a lower frequency of $\nu = 3.668$ GHz under parametric excitation. Figure~\ref{fig:lissajous}(e) shows the phase space dynamics of these two modes, which exhibit clear limit cycles albeit with larger fluctuations in comparison with the primary excitations $\psi_{11}$ and $\psi_{12}$. This suggests that the secondary excitations do not possess the same degree of coherence as the primary excitations. Nevertheless, like in the case of the primary excitations, the corresponding Lissajous curve reveals a well-defined ellipse that is indicative of mutual phase locking.

\subsection{Transient dynamics}
We now discuss the transient dynamics of the different mode populations obtained with the mode-filtering method, corresponding to the parametric excitation in Figs.~\ref{fig:psd_supercell}(b), \ref{fig:psd_supercell}(c), \ref{fig:psd_modeproj}(b), \ref{fig:psd_modeproj}(c), and \ref{fig:psd_modeproj_otherf}(b), \ref{fig:psd_modeproj_otherf}(c).
%%%
\begin{figure}
	\centering\includegraphics[width=8.5cm]{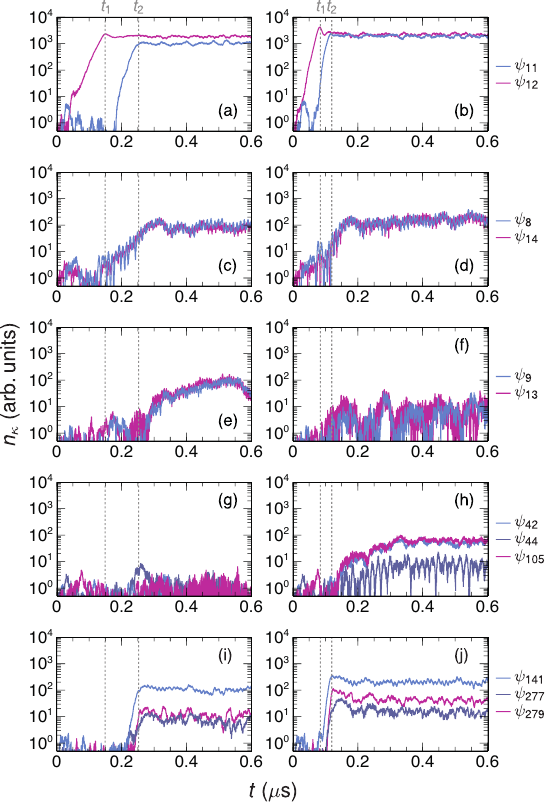}
	\caption{Transient dynamics of different mode populations $n_k$ under parametric excitation of  $b_\mathrm{rf} = 1.0$ mT (a), (c), (e), (g), (i) and $b_\mathrm{rf} = 1.5$ mT (b), (d), (f), (h), (j) at a pumping frequency of $\nu_\mathrm{rf} = 2\nu_{12} = 5.802$ GHz.}
	\label{fig:transients}
\end{figure}
%%%
We define the mode population as
\begin{equation}
n_\kappa \equiv a_{\kappa,v}^2 + \varepsilon_\kappa^2 a_{\kappa,w}^2.
\end{equation}
Consider first the growth of the modes related to the primary excitation, $\psi_{11}$ and $\psi_{12}$, the latter being the mode targeted by parallel pumping. Figure~\ref{fig:transients}(a) illustrates the transients for these modes under $b_\mathrm{rf} = 1.0$ mT. The targeted quasiuniform mode ($\psi_{12}$) is excited first, exhibiting an exponential growth in intensity from thermal levels within the first 150 ns, before reaching a steady-state level for the remainder of the 1-$\mu$s long simulation. As this level is attained at $t_1$, we observe the onset of the growth in the intensity of mode $\psi_{11}$, exhibiting a similar exponential growth before reaching in turn a steady-state level at around $t_2$. Similar behavior is observed under the stronger driving field of $b_\mathrm{rf} = 1.5$ mT as shown in Fig.~\ref{fig:transients}(b), with the primary difference being the shorter $t_1$ and $t_2$ needed to reach steady state for the populations of $\psi_{11}$ and $\psi_{12}$.

The transient dynamics of the first set of satellite peaks, $\psi_{8}$ and $\psi_{14}$, is shown in Fig.~\ref{fig:transients}(c) for $b_\mathrm{rf} = 1.0$ mT and in Fig.~\ref{fig:transients}(d) for $b_\mathrm{rf} = 1.5$ mT. Both of these modes remain at their thermal levels during initial growth of the target mode $\psi_{12}$, but grow exponentially in intensity once $\psi_{12}$ reaches steady state at $t_1$, mirroring the growth dynamics of mode $\psi_{11}$. 

In a similar way, the second set of satellite peaks, $\psi_{9}$ and $\psi_{13}$, remain at thermal levels with the onset of exponential growth triggered by $\psi_{11}$ reaching steady state, as shown in Fig.~\ref{fig:transients}(e) for $b_\mathrm{rf} = 1.0$ mT. As discussed above, the second set of satellite peaks are less perceptible under the stronger driving field of $b_\mathrm{rf} = 1.5$ mT, where population levels are an order of magnitude lower as shown in Fig.~\ref{fig:transients}(f). The satellites around $3 \nu_{12}/2$ [Fig.~\ref{fig:transients}(h)] also mirror the transient dynamics seen in Fig.~\ref{fig:transients}(e), namely that the departure from thermal levels takes place once $\psi_{11}$ reaches steady state. 

This example serves to highlight one of the strengths of the mode-filtering approach, which allows complex transient dynamics of individual modes to be quantified. In this case, parallel pumping of $\psi_{12}$ results in a cascade of secondary parametric processes, which would be difficult to ascertain from frequency-domain analyses alone.

%%
%	Parametric thresholds
%%
\section{Parallel pumping thresholds}
We turn our attention to computing thresholds for parallel pumping using mode filtering. In general, analytic expressions for the threshold field are only known for some limiting cases. For example, for plane-wave excitations in a uniformly magnetized bulk system, the threshold field for the first mode to be excited, $k$, is given by
\begin{equation}
b_\mathrm{rf,c} = \mathrm{min}\left[\frac{4 \omega_k \eta_{k}}{\gamma \omega_M \sin^2\theta_k} \right],
\label{eq:bcrit}
\end{equation}
where $\omega_M = \gamma_0 M_s$, $\eta_k$ is the Gilbert (linear) relaxation rate, and $\theta_k$ represents the propagation direction with respect to the static magnetization. We note that the $\sin^2\theta_k$ term in the denominator captures the ellipticity of the mode, which shows that the threshold tends towards infinity as $\theta_k \rightarrow 0$, i.e., for circular precession. We will revisit this point further below.

\begin{figure}
	\centering\includegraphics[width=8cm]{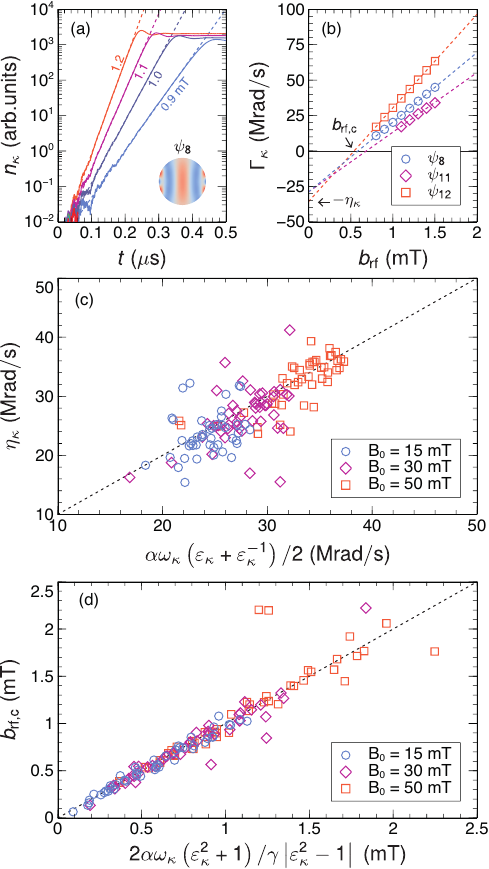}
	\caption{(a) Transient dynamics of the mode population for $\psi_8$, $n_8(t)$, for different rf field amplitudes $b_\mathrm{rf}$ at $\nu_\mathrm{rf} = 2 \nu_{8} = 5.632$ GHz. Dashed lines represent exponential fits to the initial growth before saturation. The inset shows the spatial profile of the mode. (b) Growth rate, $\Gamma_\kappa$, as a function of rf field amplitudes $b_\mathrm{rf}$ for three different modes. Dashed lines indicate linear fits, from which the critical field amplitude, $b_\mathrm{rf,c}$, and the linear (Gilbert) relaxation rate, $\eta_\kappa$ are deduced. (c) Comparison between the fitted and theoretical Gilbert damping rates for the first 50 eigenmodes under three different applied fields, $B_0$. (d) Scaling relation between the fitted critical field, theoretical Gilbert damping, mode frequency, and mode ellipticity for the data in (c).}
	\label{fig:thresholds}
\end{figure}

In contrast to bulk systems or continuous films, there are only a few (nearly-)degenerate modes in a confined system such as the in-plane magnetized disk considered here. It is therefore possible to determine the parallel pumping threshold of each mode individually by targeting it with the appropriate pump frequency, i.e., $\omega_\mathrm{rf} = 2 \omega_\kappa$. We proceed by first computing the population growth rate of a given pumped mode for different values of $b_\mathrm{rf}$, as shown in Fig.~\ref{fig:thresholds}(a) for the mode $\psi_8$.
%%%
%%%
After an initial incubation time driven by thermal fluctuations (here we take $T = 1$ K), the mode population grows exponentially until reaching a steady-state level, as discussed previously in Fig.~\ref{fig:transients}. This exponential growth can be parametrized by $n_\kappa = c \exp(\Gamma_\kappa t)$, with $\Gamma_\kappa$ representing a characteristic (mode-dependent) growth rate and $c$ a constant. The dashed lines in Fig.~\ref{fig:thresholds}(a) indicate this exponential growth, whose slope on the log-linear plot gives $\Gamma_\kappa$ directly.

The variation of the growth rate as a function of the rf field amplitude for three different modes is presented in Fig.~\ref{fig:thresholds}(b). Here, we isolate the cases in which only one mode is excited parametrically in order to avoid secondary processes as shown in Fig.~\ref{fig:transients}. For these cases, $\Gamma_\kappa$ varies linearly with $b_\mathrm{rf}$. By fitting these linear relations and extrapolating to $b_\mathrm{rf} = 0$, we can deduce two important characteristics. First, the critical field amplitude for parametric excitation, $b_\mathrm{rf,c}$, is determined by the condition for which $\Gamma_\kappa = 0$, i.e., the point at which the rate of Zeeman energy provided by the rf field that flows into the mode compensates Gilbert relaxation. Second, the intrinsic mode-dependent Gilbert relaxation rate, $\eta_\kappa$, is determined by the $y$-intercept at $b_\mathrm{rf} = 0$.

While no explicit expressions exist for the mode-dependent critical field and relaxation rate for the system considered here, we can gain some insight into their relationship with the mode frequency and ellipticity by revisiting well-known spin wave theory for an infinite medium. The usual prescription involves performing Holstein-Primakoff and Bogoliubov transformations to diagonalize the Hamiltonian (see, e.g., Refs.~\onlinecite{Sparks:1964, White:2007}), resulting in harmonic oscillator variables $c_k$ representing the spin wave eigenmodes with frequency
\begin{equation}
\omega_k = \sqrt{A_k^2 - |B_k|^2},
\label{eq:modefreq}
\end{equation}
where the coefficients $A_k$ and $B_k$ are related to the Bogoliubov transformation. For dipole-exchange spin waves in the presence of an applied external field $H_0$,  
\begin{align}
A_k   &= \left(\frac{2\gamma A}{M_s}\right) k^2 + \omega_0 + \frac{1}{2}\omega_M \sin^2\theta_k, \\
|B_k| &= \frac{1}{2}\omega_M \sin^2\theta_k,
\end{align}
where $\omega_0 = \gamma_0 H_0$. The critical field discussed above in Eq. (\ref{eq:bcrit}) pertains to this system. We recover the familiar expression form of the mode frequency
\begin{equation}
\omega_k = \sqrt{\omega_{c,k} \, (\omega_{c,k} + \omega_M)},
\end{equation}
where $\omega_{c,k} = (2\gamma A/M_s)k^2 + \omega_0$ represents the ``circular'' part of the precession (i.e., arising from the exchange and Zeeman terms) and $\omega_M$ represents the anisotropic, dipolar contribution that leads to ellipticity. Ignoring propagation losses, the Gilbert relaxation rate is given by
\begin{equation}
\eta_k = \alpha \omega_k \frac{\partial \omega_k}{\partial \omega_{c,k}} = \alpha \left(\omega_{c,k} + \frac{1}{2}\omega_M \sin^2\theta_k \right) = \alpha A_k.
\end{equation}
We can relate this to the average mode ellipticity $\varepsilon_k$ by noting that the ratio between the major and minor axes of the ellipse of precession is given by~\cite{Sparks:1964}
\begin{equation}
\varepsilon_k = \sqrt{\frac{A_k + |B_k|}{A_k - |B_k|}}.
\label{eq:ellip}
\end{equation}
By combining this with Eq.~(\ref{eq:modefreq}), we find
\begin{equation}
\eta_k = \frac{1}{2}\alpha \omega_k \left(\varepsilon_k + \varepsilon_k^{-1} \right),
\label{eq:relaxrate}
\end{equation}
which directly relates the relaxation rate with the mode frequency and ellipticity. A comparison between the predicted values of $\eta_\kappa$ given by Eq.~(\ref{eq:relaxrate}), by using the mode frequency and ellipticity found from linearization, and the fitted values based on the method in Fig.~\ref{fig:thresholds}(b) is shown in Fig.~\ref{fig:thresholds}(c). Data are given for the first 50 eigenmodes at three values of the applied field. While there is some scatter in the data, the overall trend follows the dashed line which indicates that Eq.~(\ref{eq:relaxrate}), which was motivated by arguments based on the spin wave dispersion for an infinite medium, also provides a good quantitative estimate of the relaxation rate in the in-plane magnetized disks studied.

The same line of reasoning can be employed to deduce a relationship between the threshold field for parallel pumping and the mode frequency and ellipticity. We rewrite Eq.~(\ref{eq:bcrit}) as
\begin{equation}
b_\mathrm{rf,c} = \frac{2 \omega_k \alpha A_k}{\gamma |B_k|},
\end{equation}
and substitute Eq.~(\ref{eq:ellip}) for the ellipticity, which results in the expression
\begin{equation}
b_\mathrm{rf,c} = \frac{2 \alpha \omega_k}{\gamma} \left(\frac{ \varepsilon_k^2 +1}{|\varepsilon_k^2 -1|}\right).
\label{eq:bcritellip}
\end{equation}
Figure~\ref{fig:thresholds}(d) shows a comparison between the predicted values given by Eq.~(\ref{eq:bcritellip}) and the fitted values of $b_\mathrm{rf,c}$ based on the method in Fig.~\ref{fig:thresholds}(b). As for the relaxation rate, the data are given for the first 50 eigenmodes under three values of the external field. The data show that an even better quantitative agreement is found between the predicted and fitted values of the critical field, where the majority of the data points lie along the dashed line, which indicates an equivalence between the two quantities.

%%
%	Mode generation in high power effects
%%
\section{Mode generation under high power}
As discussed in Sec.~\ref{sec:paramexc}, the mode-filtering method complements more conventional analyses based on the power spectral density by revealing features such as nonlinear frequency shifts and mutual phase locking that arise from the parametric excitation of a targeted mode. In this section, we examine a few cases of how secondary modes, which appear as satellite peaks to the primary excitation mode driven parametrically, appear as a function of the supercriticality parameter $b_\mathrm{rf}/b_\mathrm{rf,c}$.

\begin{figure}
	\centering\includegraphics[width=8cm]{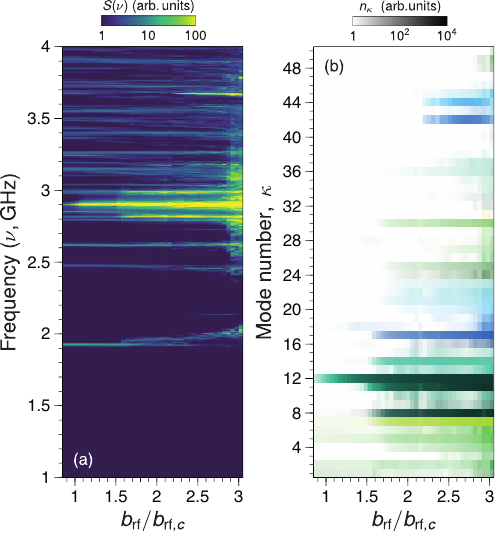}
	\caption{(a) Color map of the power spectral density, $S(\nu)$, as a function of the supercriticality, $b_\mathrm{rf}/b_\mathrm{rf,c}$ for $\nu_\mathrm{rf} = 2 \nu_{12} = 5.802$ GHz and $B_0 = 50$ mT. (b) Color map of the time-averaged population $n_\kappa$ as a function of supercriticality corresponding to the excitation in (a), where each mode is assigned a different color with an intensity corresponding to the population level on a log scale.}
	\label{fig:psdmap12}
\end{figure}

We begin by revisiting the case considered in Sec.~\ref{sec:paramexc}, where the uniform mode ($\psi_{12}$) is targeted by parallel pumping with $\nu_\mathrm{rf} = 2 \nu_{12} = 5.802$ GHz under an external field of $B_0 = 50$ mT and with $T = 300$ K. We use the mode-filtering method to compute the power spectral density and population of the modes as a function of the supercriticality, where $b_\mathrm{rf,c} = 0.51$ mT. For each value of $b_\mathrm{rf}$, time integration is performed for 500 ns and the population of each of the 50 modes is recorded at the end of simulation, while the power spectral density is computed over the entire duration of the simulation. Figure~\ref{fig:psdmap12}(a) shows a color map of the power spectral density as a function of the supercriticality.
%%%
%%%
The color map extends the analyses described in Sec.~\ref{sec:paramexc} by revealing how different satellite peaks, associated with higher-order nonlinear processes, appear as the strength of the parametric drive is increased. Below threshold, we can observe the spectrum of thermally-populated modes, as shown in Fig.~\ref{fig:psd_modeproj}(a). As the supercriticality is increased above one, the targeted mode at $\nu_{12}$ exhibits a sharp increase in intensity, while the other modes in the frequency range considered remain largely unchanged. The first set of satellite peaks appear around $b_\mathrm{rf}/b_\mathrm{rf,c} \approx 1.5$, which is accompanied by an increasing frequency splitting between the two edge modes $\psi_1$ and $\psi_2$ whose linear frequencies are 1.924 and 1.926 GHz, respectively. Around these frequencies, we can observe that a horizontal line persists up $b_\mathrm{rf}/b_\mathrm{rf,c} \approx 2.5$, indicating that one of the modes remains largely unaffected by the parametric excitation, while the other exhibits a strong positive frequency shift. At around $b_\mathrm{rf}/b_\mathrm{rf,c} \approx 2.2$, we observe the onset of a second set of satellite peaks [i.e., the lowest being around 3.67 GHz corresponding to $\psi_{42}$ and $\psi_{44}$ in Fig.~\ref{fig:psd_modeproj_otherf}(c)], followed by broadband excitations at around $b_\mathrm{rf}/b_\mathrm{rf,c} \approx 2.9$ that are suggestive of highly nonlinear and possibly chaotic dynamics.

Figure~\ref{fig:psdmap12}(b) shows a color map that encodes the response in terms of the mode populations instead. The color code indicates the population at the end of the simulation run, i.e., 500 ns after the start of the rf field excitation. This representation allows us to better identify the different modes at play, which can be masked in a spectral analysis when phenomena such as frequency shifts and phase locking take place. Besides the targeted mode, we can identify directly the modes corresponding to the secondary excitations, which become more prominent as the supercriticality is increased. Interestingly, for increasing $b_\mathrm{rf}$ we find that the onset of mode $\psi_{11}$, which subsequently becomes phase locked to the targeted $\psi_{12}$, slightly precedes with the appearance of the satellite peaks in the PSD, which are associated with the modes $\psi_{8,9}$ and $\psi_{13,14}$ as shown in Fig.~\ref{fig:psd_modeproj}.

\begin{figure}
	\centering\includegraphics[width=8cm]{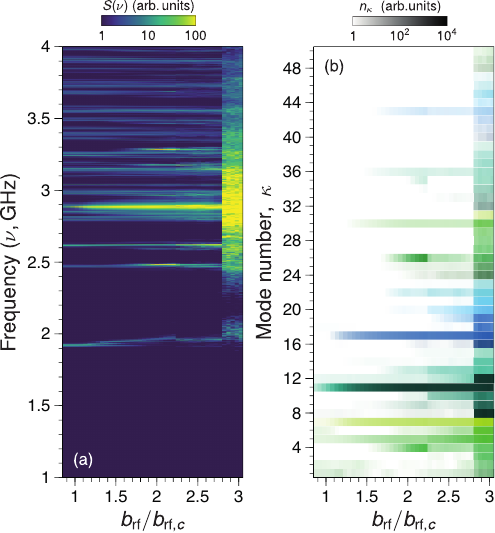}
	\caption{(a) Color map of the power spectral density, $S(\nu)$, as a function of the supercriticality, $b_\mathrm{rf}/b_\mathrm{rf,c}$ for $\nu_\mathrm{rf} = 2 \nu_{11} = 5.770$ GHz and $B_0 = 50$ mT. (b) Color map of the time-averaged population $n_\kappa$ as a function of supercriticality corresponding to the excitation in (a), where each mode is assigned a different color with an intensity corresponding to the population level on a log scale.}
	\label{fig:psdmap11}
\end{figure}

In order to better understand whether the nonlinear processes leading to the appearance of these satellite peaks result from the coupled dynamics of modes $\psi_{11}$ and $\psi_{12}$, or whether they are driven primarily by the excitation of $\psi_{11}$ alone, we present in Fig.~\ref{fig:psdmap11} the color maps for an rf excitation that targets instead $\psi_{11}$, with $\nu_\mathrm{rf} = 2 \nu_{11} = 5.770$ GHz.
%%%
%%%
Indeed, the map of the power spectral density differs qualitatively in comparison with Fig.~\ref{fig:psdmap12}(a), where secondary excitations appear at a higher supercriticality $b_\mathrm{rf}/b_\mathrm{rf,c} \approx 1.75$, with a larger frequency spacing. The transition toward the highly nonlinear regime $b_\mathrm{rf}/b_\mathrm{rf,c} \gtrsim 2.8$ also appears more abrupt, with a stronger broadband response. The marked, qualitative differences are also visible in the variation of the mode populations in Fig.~\ref{fig:psdmap11}(b) where we can identify by inspection the different modes that appear as the rf field amplitude is increased.

In general, the dynamics at high power depends very much on the targeted mode. Examples of different qualitative behavior related to other modes can be found elsewhere~\cite{SM}.

%%
%	Discussion and concluding remarks
%%
\section{Discussion and concluding remarks}
We have highlighted a number of shortcomings related to spatially resolved frequency-domain analyses, such as the supercell approach discussed here, where we encounter difficulties in identifying modes profiles at a given frequency (using the back-transformed Fourier spectra) when several modes are at play, such as in the case of mutual phase locking of $\psi_{11}$ and $\psi_{12}$ in Fig.~\ref{fig:supercell}. Such difficulties would persist even if all finite-difference cells were used in the analysis, rather than the local spatial averages used in the supercells. Mode coexistence at a given frequency results in a superposition of the spatial profiles of two (or more) eigenmodes, so one would need detailed knowledge of the relative phases between the complex mode amplitudes in order to reconstruct the mode populations from this spatial analysis.

The mode-filtering described relies on using the profiles of the different eigenmodes as spatial filters. This approach holds as long as the magnetization dynamics can be described in terms of a collection of weakly-interacting modes about the equilibrium state. However, as we have seen in Figs.~\ref{fig:psdmap12}(b) and \ref{fig:psdmap11}(b) at large supercriticality, the method returns large, non-thermal population levels across a wide range of mode indices. At these levels of parametric excitation, it is likely that the overall dynamics is in fact highly chaotic, which is reflected in the appearance of broadband noise both in the frequency domain, as seen in the power spectral density, and in the spatial domain, as characterized by $n_\kappa$.

Turning now to one of the motivations highlighted in the introduction, we discuss some perspectives on applying the mode-filtering method for tasks such as reservoir computing. The capacity to resolve mode populations directly means that outputs can be constructed from these populations, rather than on spectral signatures in the frequency domain, which, as we have shown here, can mask several modes. If we take the magnon reservoir considered by K\"{o}rber~\emph{et al.} as an example~\cite{Korber:2023pr}, mode filtering would allow machine learning for pattern recognition to be performed directly on the mode populations, rather than on frequency bins of the power spectral density. This also connects to more general concepts of reservoir computing using state variables, e.g., utilizing the Hilbert space associated with quantum systems as a computational resource~\cite{Kalfus:2022hs}. 

In summary, we have presented a mode-filtering method for micromagnetics simulations in which the spin wave mode amplitudes can be computed on the fly. The method was applied to the study of parametric excitation of spin waves in in-plane magnetized disks, where phenomena such as the transient dynamics of excited mode populations, mutual phase-locking, and frequency pulling were examined. The method complements existing techniques based on frequency-domain analysis and can be applied to studied transient processes related to nonlinear spin wave interactions.

\begin{acknowledgments}
We thank T. Srivastava, T. Devolder, and G. de Loubens for fruitful discussions. This work was supported by the European Commission's Horizon 2020 Framework Programme under Contract No. 899646 (k-Net).
\end{acknowledgments}

%%
%	References
%%
\bibliography{articles}

\end{document}